# XG-Attention-WGAN PIC: Utilizing XGboost-Attention-WGAN for Photonics Integrated Circuit Design

Mahmood Hasani, Atena Shircharandabi, Masiha Rabiei, Mobin Motaharifar, Mehrdad Ghasemi, Yaser Mike Banad, Sarah Sharif *
School of Electrical and Computer Engineering, University of Oklahoma
*Corresponding Author

*Abstract*— Photonic Integrated Circuits (PICs) are fundamental for optical computing, communication, quantum information processing, and precision sensing. However, traditional numerical simulations for designing PIC components are computationally intensive and struggle with high-dimensional parameter spaces. This paper introduces XG-Attention-WGAN PIC, a novel framework that synergistically combines Wasserstein Generative Adversarial Networks (WGANs) with eXtreme Gradient Boosting (XGBoost) to enhance parameter prediction and inverse design in PICs. We utilize Finite-Difference Time-Domain simulations to generate high-fidelity training data, which is augmented by WGAN-generated synthetic data, yielding a root mean squared error (RMSE) of 0.26089. When integrated with XGBoost, this error is reduced to 0.008. The integration of a 64-head self-attention mechanism within the WGAN generator significantly improves data quality and model efficiency over 1000 training epochs. Demonstrated on microring resonators, our approach not only achieves superior prediction accuracy and design optimization but also autonomously discovers a novel, experimentally realizable geometry with enhanced Q-factor performance. The proposed framework provides a scalable, data-driven strategy for developing high-performance PIC components, with promising implications for quantum computing and advanced optical systems.

## I. INTRODUCTION

Integrated photonics [1], integration of photonics devices on a single chip, such as waveguides, modulators, microring resonators, etc., has emerged as a growing cutting-edge field of research. Efficient and scalable photonics devices have revolutionized optical communications [2] by facilitating high-speed data transmission with lower power consumption. Additionally, integrated photonics plays a key role in sensing technology, biomedical devices, and monitoring [3-6]. As the demand for more efficient and high-performance photonic systems increases, efficient design and optimization of key components within these systems becomes increasingly crucial, and the need for innovative approaches to address these challenges becomes significantly important.

Moreover, Machine learning models, particularly deep neural networks, are characterized by tens of thousands or even millions of parameters. This abundance of parameters not only gives them great capacity to learn complex relationships, but also makes them highly susceptible to overfitting. To avoid overfitting, very large datasets are often required for effective training. However, acquiring a sufficiently large training dataset can be impractical, especially in fields like photonics. Collecting experimental data necessitates the fabrication and measurement of numerous samples, while numerical simulations (e.g., of 3D photonic crystals) are time-consuming and computationally expensive. One potential solution is to utilize synthetic data generated solely through simulation. While viable in some instances, amassing a substantial volume of purely synthetic data can still be a time-consuming and costly endeavor.

An alternative and powerful technique to achieve a sizable training set, especially when real data is limited, is data augmentation. This involves increasing the size of a dataset by generating multiple realistic variants of existing training instances. This approach is widely applied in various domains; for example, in medical imaging, data augmentation has been used to create synthetic scans to augment datasets, aiding diagnostic tools without the need for extensive real data [7, 8].

While our method could be investigated in the inverse and forward design of any photonics integrated devices, in this work, we applied our algorithm to microring resonator as an example for validating our approach. Microring resonators, a cornerstone of integrated photonics since the late 20th century [9-13], leverage Whispering-Gallery Modes (WGMs) for strong optical confinement and resonant enhancement. Their versatility spans applications in optical signal processing [14], nonlinear optics [15, 16], neuromorphic photonics [17, 18], quantum technologies [19-29], and biophotonics [5, 30-33]. Despite their significance, traditional design methods like finite-difference time-domain (FDTD) or eigenmode solvers are computationally demanding and inefficient for high-dimensional parameter spaces, especially in quantum photonics where precision is paramount [16, 34].

Recent advancements in machine learning (ML) have

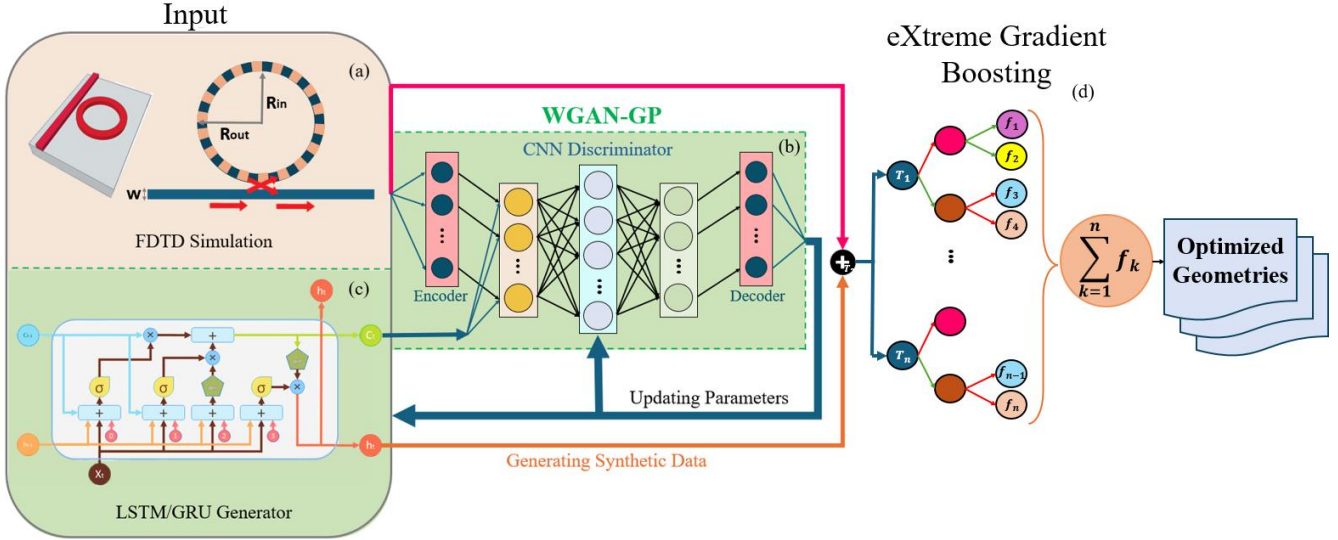

**Figure 1:** Schematic representation of the *XG-WGAN PIC* framework for photonic integrated circuit (PIC) design optimization. (a) The dataset is generated using Finite-Difference Time-Domain (FDTD) simulations, with microring resonators as a representative case study. (b) A Generative Adversarial Network (GAN) with a Convolutional Neural Network (CNN) discriminator is used to generate high-quality synthetic data. (c) A Long Short-Term Memory (LSTM) network or Gated Recurrent Units (GRUs) serves as the generator, learning to produce realistic design parameters. The generated data undergoes a condition evaluation step to ensure validity before being combined with the FDTD dataset. (d) The enriched dataset is then used to train an eXtreme Gradient Boosting (XGBoost) model, iteratively refining predictions for key parameters such as resonance wavelength and quality factor (Q). This hybrid approach enhances the accuracy and efficiency of PIC parameter prediction and inverse design, outperforming traditional simulation-based methods. In our framework, the "condition evaluation" step is a quality control process applied to the synthetic data generated by the GAN. Specifically, it involves checking that each generated sample meets predetermined physical and design constraints, such as valid ranges for geometrical parameters and expected physical behaviors in the transmission spectra. This evaluation ensures that only data samples consistent with the underlying physics of photonic integrated circuits are incorporated into the training dataset, ultimately enhancing the accuracy and reliability of the subsequent XGBoost predictions.

revolutionized predictive modeling and optimization across scientific disciplines. ML approaches for optimizing microring resonator designs, addressing the computational challenges of traditional methods have been an important area for researchers in recent years. A 2022 study [35] demonstrated the efficacy of deep learning for predicting transmittance and performing inverse design of microring resonator channel dropping filters, achieving low error rates. Additionally, a similar approach could be used in waveguide optimizations [36]. Furthermore, a recent work [37] applied Bayesian optimization to design microring resonators as quantum light sources, efficiently identifying optimal structures with high escape efficiencies and on-chip squeezing levels. Moreover, another study extended deep learning to plasmonic ring resonators, showcasing the broader applicability of machine learning in resonator design [38]. Also, Data-driven approaches have been particularly successful in tasks requiring rapid and accurate predictions, such as material property estimation of integrated circuits[39-48] and inverse photonic design [35, 38, 49-54]

Among these, eXtreme Gradient Boosting (XGBoost) [55] has gained prominence due to its efficiency, scalability, and high predictive accuracy. XGBoost has been successfully applied in diverse fields such as healthcare[56], finance [57], and material science [58], demonstrating its robustness in handling complex, high-dimensional datasets. However, despite its widespread adoption in various domains, its application in photonic device design remains largely unexplored. Compared to other ML algorithms, XGBoost's inherent advantages in efficiency, scalability, and high predictive accuracy, coupled with its robustness in handling complex, high-dimensional datasets, make it a particularly essential and effective algorithm for nanophotonic integrated design. This paper examines a simple microring resonator to illustrate how our XG-Attention-WGAN PIC algorithm overcomes these limitations, offering a scalable and efficient solution for resonator design and optimization. It is noteworthy that our work is among the first to leverage XGBoost and integrate it with a Wasserstein GAN (WGAN) [59] enhanced with Attention [60] for optical design, demonstrating its potential as a powerful optimization tool in photonics.

In particular, we aim not only to leverage the individual strengths of XGBoost and WGANs, but also to architect a collaborative system that directly advances the modeling, optimization, and geometry discovery capabilities for microring resonators. Unlike prior works that focus solely on accelerating existing designs, our goal is to identify and validate physically realizable device geometries with superior Q-factor and performance metrics. This extends beyond static prediction models to enable dynamic, data-driven generation of new structures, offering a practical and scalable pathway for inverse photonic design. This approach acknowledges both the potential and the limitations of XGBoost, especially when applied to the high-dimensional and physically constrained landscape of microring design.

However, even with its strong capabilities, XGBoost's performance can be constrained by the availability of high-quality training data, and it lacks an inherent mechanism for generative design or exploration of novel structures. To address these limitations, a WGAN is employed for data augmentation and the generation of diverse, high-fidelity design candidates.

We introduce a novel framework that integrates GANs[61, 64] with XGBoost to address the dual challenges of accurate prediction and efficient optimization of parameters for microring resonators. GANs are employed to augment limited experimental datasets by generating high-quality synthetic data, while XGBoost, a robust, efficient, and interpretable machine learning model, utilizes this enriched dataset to achieve superior prediction accuracy and computational efficiency. This synergy between WGAN-driven generative exploration and XGBoost-based predictive modeling addresses the dual challenges of accurate performance prediction and the discovery of high-impact geometries in complex photonic design spaces.

Figure 1 illustrates the overall workflow of our proposed algorithm, comprising three main stages: data generation, WGAN-based augmentation, and XGBoost-based prediction, which will be elaborated in the subsequent sections.

The remainder of this paper is organized as follows. Section 2 provides an overview of the dataset, GAN augmentation methodology, and model training in the context of novel geometry discovery. Section 3 presents the resulting optimal structure and benchmarks it against traditional simulation approaches. Section 4 discusses the implications of our findings for photonic device design. Finally, Section 5 concludes with a summary and future directions.

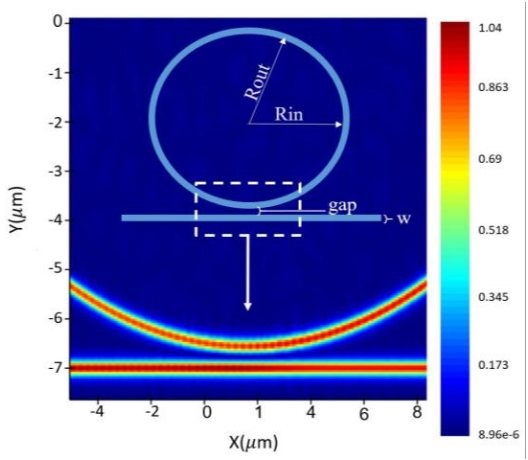

**Figure. 2**. Heatmap of the electric field distribution in the microring resonator, obtained from FDTD simulations. The high-intensity regions along the ring structure indicate strong field confinement and minimal optical leakage.

## II. SIMULATION METHODS AND DATA GENERATION

### A. Simulation-Based Input Data

To construct our initial dataset, we utilized FDTD simulations to model the optical properties of microring resonators. We employed a mode source to inject a guided wave mode into the simulation region, where a SiO2 substrate supported a Si waveguide with an overlying microring resonator. The simulations were performed under a Free Spectral Range (FSR) of 20 nm, leading to an optimal central ring radius of 18.2 μm.

The parameters varied during these simulations included the ring radius, the gap between the waveguide and the ring, and the width of both the waveguide and the ring. For simplification, the width of the waveguide and the ring were kept equal. The ring radius ranged from 17.2 μm to 19.2 μm, the coupling gap varied between 100 nm and 640 nm, and the width spanned from 200 nm to 450 nm. The structure is depicted in Fig.1(a). At the output, a transmission monitor was placed to capture resonance behavior, and the collected data served as the training foundation for our WGAN-XGBoost model [10, 62].

### B. WGAN-GP Architecture

Our framework employs the WGAN with Gradient Penalty (WGAN-GP) [63] for stable training and improved convergence as shown in Figure 1(b) and Figure 1(c). The generator and discriminator architecture are designed to capture critical spectral features and design parameter relationships.
Generator could utilize an LSTM-based architecture to model sequential dependencies in design parameters:

$$G(z, c) = f_{LSTM}(concat(z, c)) \qquad (1)$$

where the input consists of a noise vector $z \in R^{10}$ and a condition vector $c \in R^4$. The Long Short-Term Memory (LSTM) layer processes the concatenated input to learn temporal correlations in design parameter sequences, aiding in generating realistic synthetic data.

Despite the advantages of LSTM, integrating Gated Recurrent Units (GRUs) into our WGAN architecture offers several advantages over LSTM networks (see supplementary information section 1, Algorithms 2 and 3 for the pseudo code implementations of LSTM and GRU). GRUs, by design, are more streamlined, utilizing only two gates, the update and reset gates, compared to LSTM's three gates. This simplification reduces the number of parameters, leading to faster training and inference times. Moreover, GRUs have demonstrated performance comparable to LSTMs across various tasks, making them a computationally efficient alternative without compromising accuracy.

Although LSTM could be further optimized, we used GRU in this work for better and resource-optimized performance for our dataset, see figure 1(c). In future work, one can also use XLSTM [64] as generator in GAN models which can potentially provide better results.

The discriminator D is implemented as a CNN-based model to distinguish between real and generated transmission responses:

$$D(t,c) = f_{CNN}(concat(t,c)) \quad (2)$$

where the input consists of the transmission response and the condition vector. CNN employs multi-scale architecture with five convolutional layers with kernel sizes {7, 5, 3, 3, 1}y. Spectral normalization is applied to all layers to enforce the Lipschitz constraint, improving discriminator stability. See supplementary information section 1, algorithm 4 for more details of the implementation of the CNN in our PIC design.

*C. Stabilized Training Protocol of WGAN-GP*

The WGAN-GP training procedure follows a stabilized objective with gradient penalty enforcement. The discriminator loss function is given by:

$$L_D = E[D(t_{real},c)] - E[D(G(z,c),c)] + \lambda E(||\nabla_{\hat{t}} D(\hat{t},c)||_2 - 1)^2 \quad (3)$$

where $\hat{t}$ is the interpolated sample, and $\lambda$ controls the gradient penalty.

The generator loss function is:

$$L_D = -E[D(G(z,c),c)] \quad (4)$$

See the supplementary materials for section 1, Algorithm 1, for more information and details.

*D. eXtreme Gradient Boosting Architecture*

XGBoost, or Extreme Gradient Boosting, is a high-performance machine learning algorithm based on decision tree ensembles. Known for its efficiency, scalability, and robustness, XGBoost has gained significant attention for predictive modeling and optimization tasks across various scientific domains. In photonics, XGBoost provides a powerful framework for predicting key device parameters, enabling inverse design, and optimizing performance metrics of components like microring resonators[22, 35].

XGBoost builds an ensemble of decision trees, see Figure 1d, where each successive tree corrects the errors of the preceding ones. The key innovation of XGBoost lies in its formulation of the objective function, which combines a loss function with a regularization term (Ω) to prevent overfitting[22, 55]as shown in equation 5:

$$O(t) = \sum_{i=1}^{n} \mathcal{L}\left(y_i, \hat{y}_i^{(t)}\right) + \sum_{k=1}^{t} \Omega(f_k), \quad (5)$$

where $y_i$ is the true value of the $i$-th data point and $\hat{y}_i(t)$ is the predicted value after $t$ iterations, and $fk$ is the $k$-th decision tree in the ensemble. Also, $\Omega(f_k) = \gamma T + \frac{1}{2}\lambda \|\mathbf{w}\|^2$ is a regularization term based on the number of leaves $T$ and the leaf weights w, controlled by parameters $\gamma$ and $\lambda$.

The optimization process in XGBoost uses second-order Taylor expansion for the loss function, enabling faster convergence [65]:

$$\mathcal{L}(t) \approx \sum_{i=1}^{n} \left[g_i \Delta f(x_i) + \frac{1}{2} h_i \left(\Delta f(x_i)\right)^2\right] \quad (6)$$

where $gi$ and $hi$ are the first and second-order gradients of the loss function concerning the predictions.

Figure 1d presents the workflow of the XGBoost-based predictive modeling process as integrated within the WGAN PIC framework. The diagram outlines each step starting from feature extraction from the augmented dataset, proceeding through train-test splitting, and advancing to the construction and training of decision trees via gradient boosting (see supplementary information section 1, Algorithm 5 for more details on the XGBoost implementation). The ensemble model is then refined through an iterative feedback loop that adjusts the weights based on prediction errors. This structured approach ensures that the XGBoost model effectively captures the complex, nonlinear relationships between the input design parameters and the key photonic performance metrics, thereby enabling rapid and accurate device optimization.

III. FEATURE ENGINEERING FOR PHOTONICS DESIGN

Raw transmission spectra acquired from ring resonator measurements are inherently high-dimensional and frequently contaminated by measurement noise, baseline drifts, and strong inter-wavelength correlations. Direct ingestion of such data into machine learning models can lead to poor convergence, overfitting, and limited interpretability. By applying a systematic preprocessing pipeline, including denoising, baseline correction, and normalization, we can overcome challenges such as convergence, etc. In our method, we preprocess the transmission data obtained from ring resonators to extract meaningful features. These features enhance the predictive capability of machine learning models, including GANs and gradient-boosting algorithms such as XGBoost. The feature engineering involves applying various technical indicators inspired by financial time series analysis to capture short-term and long-term variations in transmission characteristics.

The feature engineering process consists of the following sequential steps: Data Loading and Sorting. In this method, the dataset, stored in CSV format, is loaded into a Pandas DataFrame. The data is sorted based on the wavelength column to ensure chronological order. Another method is Feature Extraction using Technical Indicators. In this method, we use Exponential Moving Average (EMA), Moving Average Convergence Divergence (MACD), Relative Strength Index (RSI), Bollinger Bands, and Relative Strength Value (RSV) to capture essential transmission characteristics.

The EMA smooths short-term fluctuations in transmission and highlights trends. We compute EMAs with window sizes of 2, 3, and 5 points using:

$$EMA_t = \alpha \cdot X_t + (1-\alpha) \cdot EMA_{t-1} \quad (7)$$

where $\alpha = \frac{2}{n+1}$ and $X_t$ represents the transmission at time t.

MACD is computed as the difference between short-term and long-term EMAs. It provides insights into trend direction and momentum shifts. We calculate MACD using different parameter sets:

$$MACD = EMA_{short} - EMA_{long} \quad (8)$$

where short and long refer to different EMA window lengths (e.g., 2-3-5 and 3-5-7). RSI measures the speed and changes of transmission intensity variations. It is computed as:

$$RSI = 100 - \left(\frac{100}{1 + RS}\right) \quad (9)$$

where $RS$ is the ratio of average gain to average loss. We calculate RSI for 2, 3, and 5-point windows to capture varying levels of sensitivity.

Bollinger Bands provide an envelope around transmission variations based on standard deviation (σ) from the moving average:

$$Upper\ Band = SMA + k \cdot \sigma \quad (10)$$
$$Lower\ Band = SMA - k \cdot \sigma \quad (11)$$

where k is set to 2. We compute bands for 2, 3, and 5-point windows. RSV normalizes the transmission relative to its local minimum and maximum:

$$RSV = \frac{X_t - X_{min}}{X_{max} - X_{min}} \quad (12)$$

where $X_{min}$ and $X_{max}$ are the minimum and maximum transmission values within a given window. We compute RSV for 2, 3, and 5-point intervals. By utilizing these methods, we generated 17 more features, and as you can see in the next section, integrating all works together, the algorithm works at its best performance.

### IV. Results and Discussion

In this section, we want to discuss the results from each machine learning algorithm for the photonics integrated circuit design step by step. We first dive into the WGAN for PIC optimization and data augmentations and compare different hyperparameters. Then we evaluate the optimizations on the WGAN, such as adding attention and VAE. Additionally, using advanced preprocessing methods, we optimized the WGAN model to the best of our knowledge. Finally, we integrate all the methods mentioned with XGBoost for the final results.

*A. WGAN for Predictive Modeling and Design Optimization in Photonics*

As discussed earlier, we want to conduct dataset augmentation with WGANs as the second step in our hybrid ML algorithm, see Figure 1b and Figure 1c, WGAN is employed to generate synthetic data samples that expand the diversity and size of the available datasets. For example, given a limited dataset of microring resonators characterized by Q and $\lambda_{res}$, a WGAN can learn the underlying data distribution and synthesize additional samples. This augmentation is particularly valuable in addressing sparsity in experimental data

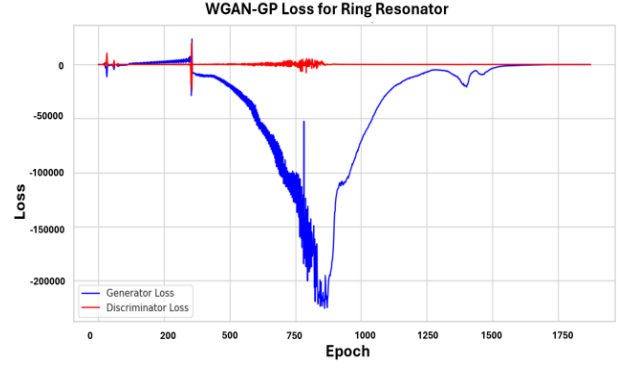

**Figure 3.** Generator loss (depicted in blue) and the discriminator loss (in black) are plotted over the first 1000 epochs. Initially, the generator loss starts at a high value and decreases rapidly within the first 100 epochs, indicating that the generator is learning to produce more realistic samples. However, fluctuations and instability are observed in the generator loss, especially around epochs 100 to 300. This may be due to mode collapse or oscillations in the adversarial training dynamics. The discriminator loss remains close to zero, suggesting that the gradient penalty stabilizes the training, though occasional sharp dips could indicate moments of instability in weight updates. Around epoch 600, a significant increase in generator loss is observed, peaking around epoch 800, before declining again. This suggests that the generator struggled to produce realistic samples during this phase. This behavior may indicate an imbalance between the generator and discriminator learning rates or improper weight clipping in the WGAN framework.

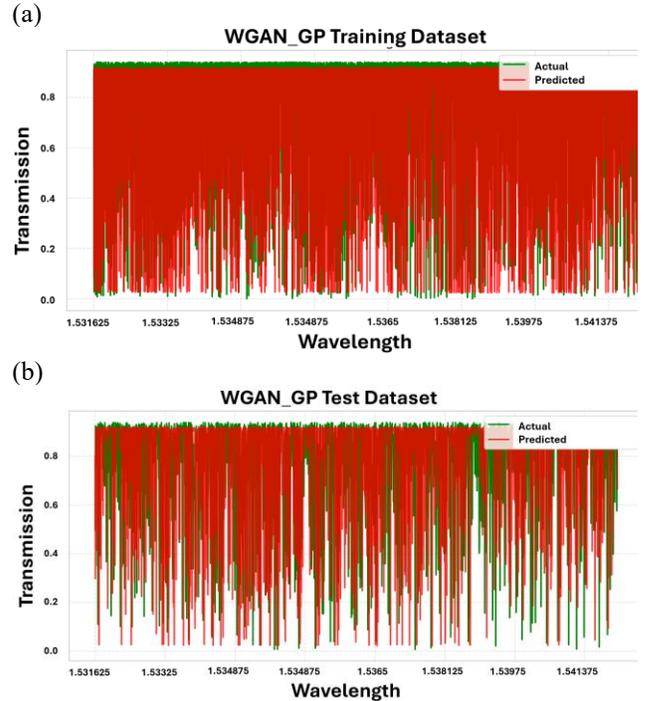

**Figure 4.** Comparison of actual and predicted transmission data using the WGAN-GP model after 1000 training epochs. The actual transmission is represented by the green curve, while the predicted transmission is shown in yellow. At this stage, the model has captured the general structure of the transmission data but still exhibits some discrepancies, particularly in lower transmission regions, suggesting the need for further training to enhance accuracy. In this plot, the learning rate is 0.000115.

and exploring design spaces that are computationally expensive to simulate[7, 66].

In step one, we illustrate the performance of WGAN on the ring resonator. Next, we trained a WGAN-GP to generate synthetic transmission spectra and corresponding geometrical parameters, including ring radius, coupling gap, and waveguide width. The generated data was evaluated for physical plausibility and consistency with known optical behaviors and then integrated with the original FDTD dataset to enhance the training of downstream predictive models. It is noteworthy that this section analyses are without the encoder and decoder.

To evaluate the performance of our Wasserstein Generative Adversarial Network with Gradient Penalty (WGAN-GP) model in generating transmission spectra for micro-ring resonators, we analyzed both the training dynamics and the quality of the generated data compared to real samples.

The loss curves, shown in Figure 3, demonstrate stable convergence of both the generator and discriminator losses after an initial fluctuation. The generator loss exhibits a gradual upward trend, while the discriminator loss remains near zero, indicating that the model has reached an equilibrium where the generator produces samples that the discriminator finds challenging to distinguish from real data. The stability of the WGAN-GP training process, particularly after 1500 epochs, confirms the effectiveness of the gradient penalty in mitigating mode collapse and ensuring smooth learning. The loss curves demonstrate that, after an initial period of fluctuation, both networks achieve stable convergence. The discriminator's loss, remaining close to zero, indicates that the generator is successfully producing synthetic transmission spectra that are quite perfectly aligned with real data. Concurrently, the gradual upward trend in the generator loss reflects ongoing improvements in the fidelity of the samples generated. These results confirm that the incorporation of the gradient penalty effectively mitigates issues like mode collapse and facilitates a robust training process.

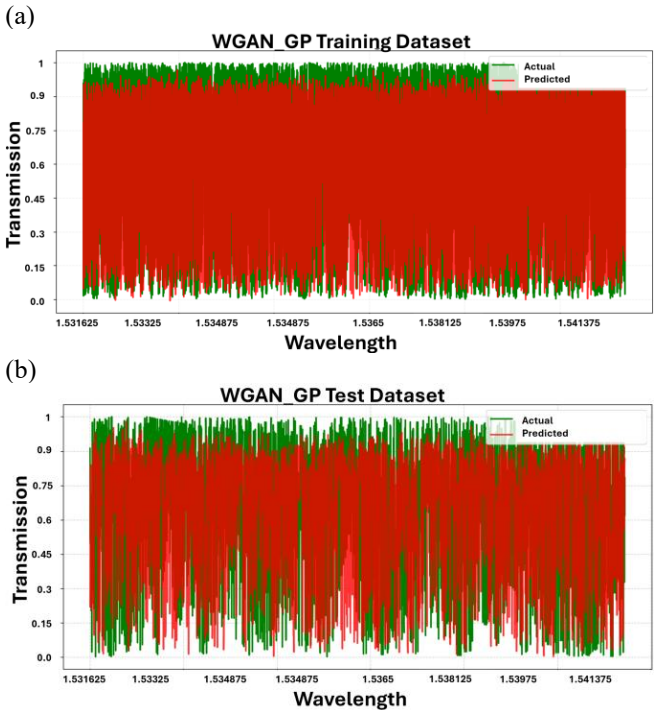

**Figure 5.** Performance of the WGAN-GP model in predicting transmission data after 2000 training epochs. The green curve represents the actual transmission, while the yellow region corresponds to the predicted values. Compared to the 1000-epoch case, the model demonstrates less alignment with the actual data, indicating enhanced learning of the transmission function. The learning rate in these plots is 0.001.

While some discrepancies exist, the generated samples largely fall within the expected range, suggesting that the GAN has learned the underlying mapping from input features to transmission values. In addition, Figure 4 shows the comparison between the generated and real transmission responses between the training dataset, Figure 4(a), and the test dataset, Figure 4(b). The results show that after training on the dataset, the test split of our data could be predicted by our model.

The result of this section confirms the stability and scarcity of our data generation with WGAN. This result illustrates that the GAN can produce high-fidelity synthetic data that closely follows the underlying physics of the resonator design.

Moreover, GANs need lower computational costs and time requirements, indicating computational efficiency of GANs by reducing their reliance on full-wave simulations. Finally, exploration of novel designs using GANs facilitates the discovery of new device configurations by extrapolating beyond the training dataset, Figure 1(b) and Figure 1(c). Additionally, GANs play a critical role in inverse design by generating data that corresponds to desired performance metrics. For instance, given a target Q or $\lambda$res, GANs can suggest plausible geometric configurations that meet these requirements. This capability significantly accelerates the design process compared to traditional trial-and-error approaches.

Despite their advantages, GANs face challenges such as mode collapse, where the generator produces a limited variety of outputs. Techniques like WGANs and spectral normalization can address these issues by stabilizing training and ensuring diversity. Furthermore, integrating GANs with other generative frameworks, such as Variational Autoencoders (VAEs), may enhance their applicability to complex photonic systems[35, 66]. GANs could be beneficial in the design and optimization of photonic devices. By synthesizing high-quality data and enabling efficient inverse design, they complement traditional simulation methods and empower machine learning models like XGBoost to achieve superior predictive performance. This integration marks a significant step toward scalable, data-driven methodologies in photonic engineering.[22, 35, 55, 65]

In Figure 5, we compare the synthetic transmission spectra generated by the GAN with the real spectra obtained from our FDTD simulations for different hyperparameters. We set the learning rate of the WGAN to 0.001 for 2000 epochs. While we increased the number of epochs alignment between the two datasets decreased due to the inability to achieve relaxation in the training procedure. Thus, choosing the right hyperparameters plays a key role in the optimization process and the overall design.

*B. Variational Autoencoder (VAE) for Latent Space*

By using VAE-WGAN, we ensure that the generator produces physically meaningful photonic device structures while maintaining a well-organized latent representation. Our VAE consists of an encoder, latent space representation, and decoder, designed to model complex photonic design distributions.

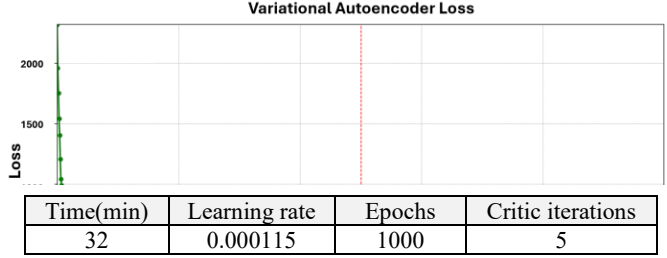

| Time(min) | Learning rate | Epochs | Critic iterations |
|---|---|---|---|
| 32 | 0.000115 | 1000 | 5 |

**Table I.** parameters of WGAN-GP for data augmenting. The model trained on 4090 GPU system.

**Figure 6.** This figure illustrates the training loss evolution of the Variational Autoencoder (VAE) over 500 epochs. The loss initially exhibits a sharp decline, indicating rapid convergence in the early stages of training. The oscillations in the mid-training phase reflect the model's balancing act between reconstruction loss (ensuring accurate recovery of photonic designs) and Kullback-Leibler (KL) divergence (regularizing the latent space). After approximately 300 epochs, the loss stabilizes, suggesting that the VAE has successfully learned a structured and smooth latent representation. This structured latent space is crucial for improving the stability of the GAN, enhancing inverse design capabilities, and ensuring meaningful interpolation between generated photonic device configurations.

The encoder consists of multiple fully connected layers with ReLU activations, progressively reducing dimensionality. It maps the input photonic device parameters to a latent distribution parameterized by mean (μ) and log-variance ($log\sigma^2$).

The latent representation is extracted using the layers:

$$\mu = f_\mu(Encoder(x)), \quad log\sigma2 = f_\sigma(Encoder(x)) \qquad (13)$$

Instead of directly sampling from the latent distribution, we use the reparameterization trick to ensure differentiability:

$$z = \mu + \sigma.\epsilon, \quad \epsilon \sim N(0,1) \qquad (14)$$

This allows the model to learn a meaningful, smooth latent space, ensuring that small perturbations in z lead to gradual changes in the output photonic design.

The decoder mirrors the encoder but reconstructs the original photonic structure from z. The final layer uses Sigmoid activation, ensuring valid normalized outputs. The decoder ensures smooth mapping from the latent space back to the physical design space.

The role of VAE in training stability and inverse design could be presented in various elements. First, the VAE improves GAN's Training Stability, which provides a structured prior distribution, ensuring that the GAN does not suffer from unstable training. This prevents collapse and leads to more diverse, physically realistic synthetic photonic designs. It also provides better inverse design capabilities by constraining the latent space, and inverse mapping from output to input parameters becomes more efficient. Small movements in latent space correspond to meaningful changes in photonic design parameters. Additionally, it helps with denoising and regularization. The encoder-decoder structure acts as a regularizer, moving irrelevant variations and noise. This is particularly useful in handling experimental noise in photonic circuit fabrication.

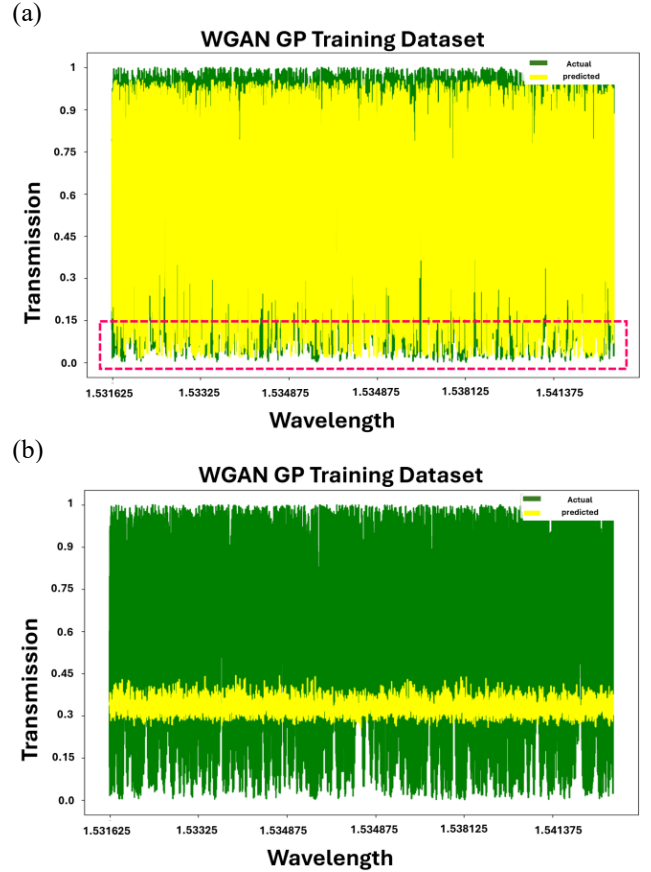

**Figure 7.** Comparison of actual and predicted transmission spectra using WGAN-GP with attention mechanisms. The first plot shows results using WGAN-GP with 128 attention heads, while the second plot shows results using WGAN-GP with 64 attention heads. The green curve represents the actual transmission, while the yellow curve represents the predicted transmission. The results demonstrate the impact of attention mechanisms on the model's ability to capture transmission characteristics. The RMSE on the dataset with 64 attention heads is 0.26, and the RMSE of the 128 attention heads is 1.23.

Figure 6 represents the VAE loss curve during training. The sharp initial drop in loss suggests effective convergence in the early training stages. The oscillations indicate that the VAE is balancing reconstruction loss (data fidelity) and Kullback-Leibler (KL) divergence (latent space regularization). The smooth stabilization after ~300 epochs confirms that the VAE has successfully learned a meaningful latent representation.

*C. WGAN with Attention*

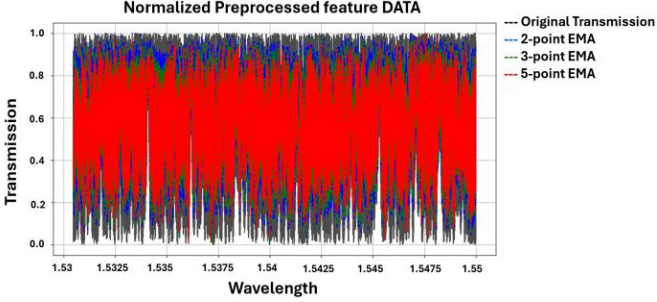

**Figure 8.** Normalized preprocessed features applied to transmission data. The plot illustrates the original transmission data (black) alongside the exponential moving averages (EMAs) computed with window sizes of 2 (blue dashed), 3 (green dashed), and 5 (red dashed). The x-axis represents the wavelength in micrometers, while the y-axis shows the normalized values of transmission and its derived features. The EMA smoothing technique reduces high-frequency noise and enhances trend visibility in the optical response of the ring resonator. The normalization ensures that all features are scaled consistently, facilitating further analysis and modeling. The presence of dense oscillations in the raw transmission data highlights the intricate spectral behavior of the resonator, while the progressively smoothed EMA curves provide different levels of trend extraction, preserving essential variations while reducing noise.

| Time(s) | Number of estimators | Learning rate | Initial random state | Max depth |
|---|---|---|---|---|
| 42 | 10000 | 0.1 | 42 | 60 |

**Table II.** Parameters of the XGBoost model.

Our framework employs the Wasserstein Generative Adversarial Network with Gradient Penalty (WGAN-GP) [63] for stable training and high-fidelity synthetic data generation. To further enhance the generator's ability to model complex relationships in photonic design parameters, we integrate an Attention mechanism into the generator architecture, replacing the previously described Gated Recurrent Unit (GRU)-based design. This Attention-enhanced WGAN-GP, referred to as Attention-WGAN-GP, leverages the self-attention mechanism [60] to capture long-range dependencies and prioritize critical design features, improving the quality of synthetic data for microring resonator optimization.

For instance, we can consider that the generator in our Attention-WGAN-GP framework is designed to produce realistic design parameters, such as ring radius, coupling gaps, and waveguide width, conditioned on a noise vector ($z \in \mathcal{R}^8$) and condition vector ($c \in \mathcal{R}^6$). Unlike the GRU-based generator, which models sequential dependencies, the Attention-enhanced generator incorporates a multi-head self-attention mechanism to focus on relevant input features and their interactions across the parameter space.

The generator architecture is defined as follows:

$$G(z,c) = f_{Attetntion}(concat(z,c)) \quad (15)$$

The input consists of a concatenated vector of the noise vector (z) and condition vector (c), forming a feature. This vector is passed through a fully connected layer to project it into a higher dimension for subsequent processing.

| Paper | Year | Method | RMSE |
|---|---|---|---|
| High-efficiency reinforcement learning with hybrid architecture photonic integrated circuit | 2024 | Reinforcement Learning with PIC | 0.0012 |
| Benchmarking deep learning-based models on nanophotonic inverse design problems | 2022 | Deep Learning (various) | 0.0829 (GANs, highest) |
| Reflective microring-resonator-based microwave photonic sensor incorporating a self-attention assisted convolutional neural network | 2024 | Self-Attention CNN | 0.026 |
| Deep learning for the design of nano-photonic structures | 2023 | Deep Learning (ANNs) | 0.015 |
| XG-Attention-WGAN PIC: Utilizing XGboostAttention-WGAN for Photonics Integrated Circuit Design | 2025 | WGAN-GP with Attention, XGBoost | 0.008 |

**Table III.** Shows the comparison of previous works with this paper. This paper illustrates a novel ML algorithm for photonics design could be enhanced by hybrid ML algorithms. While the RL method outperformed our method, our method consumed only more than 20 minutes for 25000 datapoints.

The projected features are processed by a multi-head self-attention layer, inspired by the Transformer architecture [60]. The attention mechanism computes attention scores as:

$$Attention(Q,K,V) = softmax\left(\frac{QK^T}{d_k}\right)V \quad (16)$$

where Q, K, and V are query, key, and value matrices derived from the input features, and $d_k$ is the dimension of the keys. We use 16 and 64 attention heads. This allows the generator to focus on critical geometric and material properties that influence the transmission.

Consequently, the output of the attention layer is processed by a feed-forward neural network (FFN) with two linear transformations and a ReLU activation. The final layer maps the processed features to the desired output space, producing synthetic design parameters that adhere to the physical constraints of microring resonators.

This architecture is more expressive than the GRU-based generator, as it captures both local and global dependencies in the design parameter space when using 64 heads. For more details, you can see the supplementary materials section 2 for the behavior of the loss of the generator and its dynamics. The attention mechanism enables the generator to prioritize features that significantly impact the transmission spectra, such as the coupling gap, which is critical for achieving high Q-factors.

*D. Data-Driven Discovery of Optimized Microring Geometries via Integrating WGAN-Attention and XGBoost*

The integration of XGBoost with the Attention-enhanced Wasserstein Generative Adversarial Network with Gradient

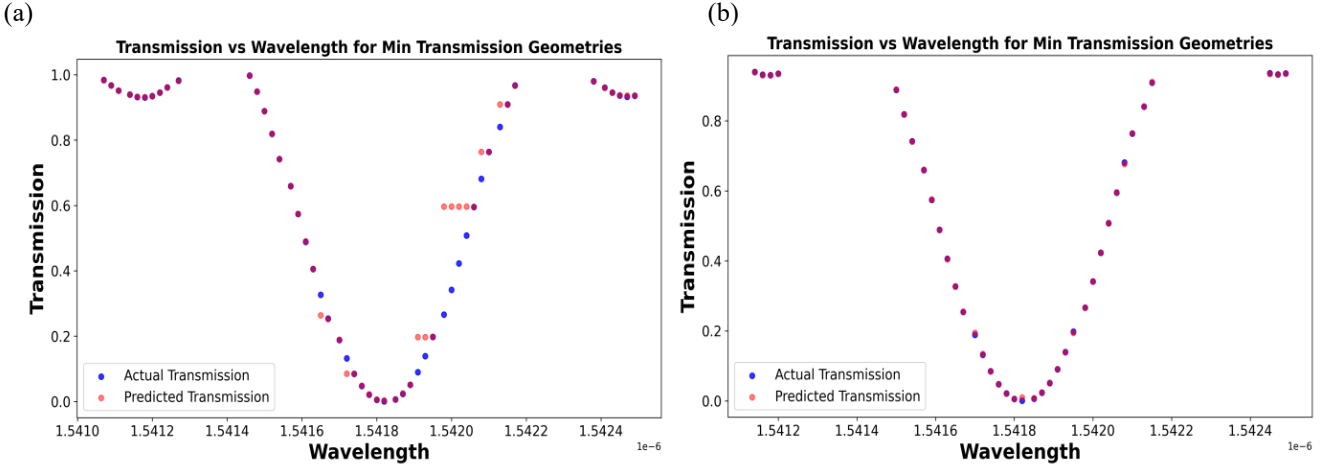

**Figure 9**. The accurate prediction of performance metrics such as the quality factor ($Q$) and resonance wavelength ($\lambda_{res}$) is essential for the design and optimization of microring resonators. XGBoost excels in these tasks by effectively modeling the complex, nonlinear relationships between input features (e.g., geometrical dimensions, material properties) and output parameters. In (a) we just use a simple XGBoost with Grid search, and in (b) we used preprocessing in addition to the WGAN-Attention as generative model play a crucial role for generating optimized geometry.

Penalty (WGAN-Attention) significantly enhances the design and optimization of microring resonators by improving the prediction of key performance metrics, such as quality factor (Q) and resonance wavelength. This section presents experimental results, demonstrating the effectiveness of our XG-WGAN PIC framework through comparisons shown in Figures 9 and 10.

Figure 9 compares the predictive performance of two approaches: (a) a simple XGBoost model with grid search, and (b) an XGBoost model enhanced with preprocessing and WGAN-Attention-generated synthetic data. In Figure 9(a), the simple XGBoost model achieves reasonable accuracy in predicting (Q) and $\lambda_{res}$, but it struggles with capturing the complex nonlinear relationships between geometric parameters (e.g., coupling gap, ring radius) and performance metrics, leading to higher prediction errors. In contrast, Figure 9(b) shows that the WGAN-Attention-augmented XGBoost model, combined with preprocessing techniques like Exponential Moving Average (EMA) smoothing (Section VII, Figure 8), significantly reduces prediction errors. The synthetic data generated by WGAN-Attention enriches the training dataset, enabling XGBoost to better model the intricate dependencies in the photonic design space. This results in more accurate predictions of ( Q ) (e.g., achieving a predicted ( Q ) close to the experimental value of 12,141) and ($\lambda_{res}$), as validated by FDTD simulations [35].

Figure 10 further illustrates the improvement in predicting transmission spectra. In Figure 10(a), the simple XGBoost model predicts transmission spectra with noticeable discrepancies compared to the actual spectra from FDTD simulations, particularly in regions with sharp resonance peaks. These errors stem from the limited diversity of the training dataset. In Figure 10(b), the XGBoost model trained on WGAN-Attention-generated synthetic data shows a much closer alignment between predicted and actual transmission spectra. The Attention mechanism in the generator (Section VII, Equation (16)) prioritizes critical design parameters, such as the coupling gap, ensuring that the synthetic data captures the underlying physics of microring resonators. This high-fidelity synthetic data enables XGBoost to produce transmission spectra that closely match the real data, as evidenced by the improved overlap in resonance peak positions and amplitudes after 2000 training epochs (Figure 5).

The integration of WGAN-Attention with XGBoost offers several key benefits for photonic design. The enriched dataset from WGAN-Attention allows XGBoost to capture complex relationships, reducing errors in predicting (Q), and transmission spectra, as shown in Figures 9 and 10. Moreover, by generating synthetic data that corresponds to desired performance metrics (e.g., high (Q)), the framework accelerates the identification of optimal design parameters, such as the coupling gap of 190 nm and ring radius of 18.2 µm used in our optimized design (Section II.A). The use of synthetic data reduces reliance on time-consuming FDTD simulations, lowering computational costs while maintaining high accuracy [66]. The framework's ability to generate diverse synthetic samples supports the design of more complex PIC components, addressing the needs of quantum photonics applications (Section I).

The training stability of the WGAN-Attention model, as shown in Figure S1 in Supplementary Information, further supports these results. The loss curves indicate stable convergence, with the Attention mechanism mitigating mode collapse and ensuring diverse synthetic samples. This stability translates to a robust dataset for XGBoost, enhancing its predictive performance. The experimental results validate the effectiveness of integrating XGBoost with WGAN-Attention in the XG-WGAN PIC framework. The improved predictions of (Q), ( $\lambda_{res}$), and transmission spectra (Figures 9 and 10) demonstrate the framework's potential to streamline microring resonator design and optimization, see Table III. These advancements pave the way for scalable, data-driven approaches in photonic engineering, with applications in quantum computing and optical communications [7].

### E. Computational Cost

In this work, as previously mentioned, we began by conducting FDTD simulations to generate real data for our machine learning (ML) design experiments. Following this, we employed both WGAN and XGBoost models for predicting transmittance spectra, and integrated them to enhance and validate overall performance. In the WGAN framework, we incorporated a Variational Autoencoder (VAE) to enable more efficient training in the latent space. Our FDTD simulations produced over 25,000 data points. These simulations were distributed across three systems, each with a single-thread CPU and 16 GB of RAM, and required approximately 13 hours to complete. As discussed in Section II.A, we varied key geometrical parameters of the microring resonator, including width, inner radius, outer radius, and gap.

For the WGAN model, training was conducted using a batch size of 128 for 1000 epochs on a system equipped with an NVIDIA RTX 4090 GPU, which yielded the results shown in Figures 4, 5, and 6. This highlights the advantage of generative models—not only do they help enrich datasets when experimental data is limited, but they are also significantly less time-consuming compared to FDTD simulations. It is noteworthy to mention that we tested the batch size of 256, and with the batch size of 256, we also reached it in less than 20 minutes with quite similar results.

Furthermore, we extended our study by integrating the XGBoost model with WGAN-generated data. By injecting 5,000 synthetic data points into the real dataset and shuffling the combined data, we observed a noticeable improvement in XGBoost's performance. This enhancement, which required less than a minute of computation, is illustrated in Figure 10.

## V. CONCLUSION

This paper presented XG-WGAN PIC, a novel data-driven framework that synergistically integrates Generative Adversarial Networks with XGBoost for the design and optimization of photonic integrated circuits. By employing FDTD simulations to generate high-fidelity baseline data and augmenting it with high-quality synthetic samples produced by a WGAN-GP, the framework overcomes the limitations of traditional, computationally intensive simulation methods. Our results, demonstrated on microring resonators, confirm that the proposed approach not only achieves superior prediction accuracy for key performance metrics such as the quality factor and resonance wavelength but also significantly accelerates the inverse design process. Furthermore, the interpretability of the XGBoost model provides valuable insights into the influence of various design parameters, thereby facilitating efficient optimization. While RL PIC outperforms our method in RMSE, our method is less time-consuming and generates new geometries that are practical and could be enhanced with RL too. Overall, XG-WGAN PIC offers a scalable and robust methodology for advancing photonic device design, with promising implications for quantum computing, secure communications, and advanced optical systems. Future work will extend this framework to a broader range of PIC components and more complex photonic architectures.

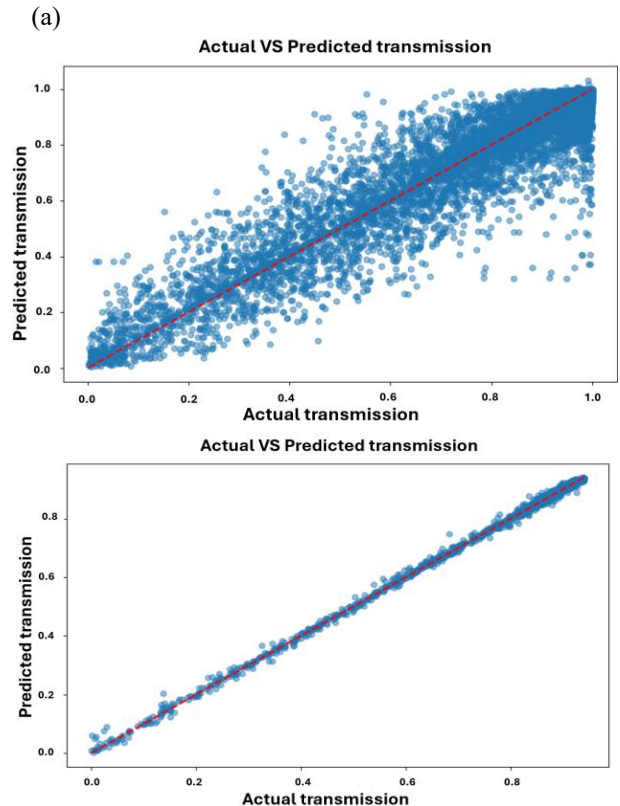

**Figure 10.** Comparison of actual and predicted transmission spectra for microring resonators. (a) Predictions using a simple XGBoost model with grid search, showing discrepancies in resonance peak alignment compared to FDTD simulation data. (b) Predictions using XGBoost trained on synthetic data from the WGAN-Attention model, demonstrating improved alignment with actual spectra, particularly in capturing sharp resonance features, due to the high-fidelity synthetic data generated by the Attention mechanism.

## VI. APPENDIX

### A. Theoretical Framework

The free spectral range (FSR) is a key parameter in microring resonator design, representing the spacing between consecutive resonances[35]. It is defined as:

$$FSR = \frac{\lambda^2}{2\pi R n_g} \quad \text{A.1}$$

where $\lambda$ is the operating wavelength, $R$ is the microring radius, and $n_g$ is the group refractive index. In this study, the operating wavelength is set to 1550 nm, and the refractive index of silicon is n = 3.4792, obtained from experimental measurements and verified using the data in [67]. Given a target FSR of 21 nm, the radius of the micro ring resonator is calculated as $R = 18.2 \ \mu m$.

The group refractive index $n_g$ accounts for the dispersive properties of the waveguide material and is calculated using:

$$n_g = n_{eff}(\lambda) - \lambda \frac{dn_{eff}}{d\lambda} \quad \text{A.2}$$

where $n_{eff}(\lambda)$ is the effective refractive index at the wavelength $\lambda$, and $\partial n_{eff}$ represents the wavelength dispersion of $n_{eff}$. For this design, $n_{eff}(\lambda) = 3.4792$, and $\frac{dn_{eff}}{d\lambda}$ found by simulation data, resulting $n_g = 3.5997$.

Coupling-induced frequency shifts (CIFS) are a critical factor in resonator design and may result from fabrication imperfections or variations in the waveguide-resonator interaction [7]. These shifts are corrected using the following adjustment formula:

$$dR = \frac{\lambda}{4\pi n_{eff}} d\phi,$$

where $n_{eff}$ is the effective refractive index, and $d\phi$ represents the phase shift introduced by coupling.

*B. Numerical and Experimental Analysis*

The quality factor ($Q_L$) describes the sharpness of the resonance and is defined as:

$$Q_L = \frac{\lambda_{res}}{FWHM} \qquad \text{B.1}$$

where $\lambda_{res}$ is the resonance wavelength, and FWHM is the full width at half maximum of the resonance peak.

Additionally, the interaction between the microring resonator and the waveguide is described using coupling coefficients. The power coupling model is expressed as:

$$\begin{pmatrix} E_{t1} \\ E_{t2} \end{pmatrix} = e^{i\phi_t} \begin{pmatrix} t & i\kappa \\ i\kappa & t \end{pmatrix} \begin{pmatrix} E_{i1} \\ E_{i2} \end{pmatrix} \qquad \text{B.2}$$

where $|t|^2 + |\kappa|^2 = 1$, and $t^2$ is derived from simulations, representing the power splitting ratios.

Finally, in modeling the transmission spectrum of the microring resonator, the Lumerical FDTD Solutions was used. The refractive index of the silicon ($n = 3.4792$) at a wavelength of 1550 nm was utilized. Coupled gap values and radii were varied to match the spectral response.

In the context of photonic systems, where predicting and optimizing parameters like the quality factor (Q) and resonance wavelength are critical, GAN offers a robust framework for augmenting datasets, reducing computational costs, and exploring novel device configurations. GANs have been employed to generate artificial training data, Figure 1(b), which is particularly useful in situations with imbalanced datasets or where data contains sensitive information. This synthetic data generation can aid in tasks such as inverse design and optimization of photonic devices, enabling the synthesis of high-quality data and facilitating inverse design tasks. Despite their impressive capabilities, the computational complexity and memory demands of GANs present significant challenges for traditional electronic accelerators. However, advancements in photonic computing could offer potential solutions to these challenges, providing faster and more efficient processing capabilities for training GANs and other deep learning models.

Following the determination of the optimal ring radius of 18.2 μm (identified through the WGAN-XGBoost framework, as will be detailed later), we proceeded with the precise design of the microring resonator to ensure a maximized quality factor (Q). The final optimized structure was defined with a coupling gap of 190 nm, an inner radius of 18.2 μm, and a waveguide width of 220 nm.

Using these optimized parameters, the resonator achieved an exceptionally high-quality factor of Q = 12,141, significantly outperforming comparable designs reported in previous studies. The obtained Q-factor is indicative of reduced optical losses and enhanced resonance sharpness, which is crucial for high-performance photonic applications. This result validates the effectiveness of our combined GAN-XGBoost framework in optimizing microring resonator designs, achieving superior performance beyond conventional simulation methods. A.3

The final microring structure is illustrated in Fig. 2, where the optimized parameters and their influence on resonance behavior are depicted. To further analyze the electromagnetic field confinement within the microring resonator, a heatmap of the electric field intensity was generated based on FDTD simulations. Fig. 2 illustrates the spatial distribution of the electric field intensity across the microring resonator, where red regions correspond to high field intensity, and blue regions indicate low intensity. This heatmap provides significant insights into the resonance conditions and the mode confinement efficiency of the structure. It clearly shows strong field localization along the ring edges, with a noticeable concentration near the coupling region. The field distribution confirms that the optimized design supports a high-Q resonance mode with minimal optical leakage, which aligns well with the previously obtained Q-factor.